\documentclass{elsart}
\usepackage{epsfig}
\newcommand{\bm}[1]{ \mbox{\boldmath $#1$}  }
\begin{document}
\begin{frontmatter}

\title{Origin of Borromean systems}

\author{E. Garrido} 
\address{Instituto de Estructura de la Materia, CSIC, Serrano 123, E-28006
Madrid, Spain}
\author{D.V. Fedorov and A.S. Jensen}
\address{Department of Physics and Astronomy, University of Aarhus,
DK-8000 Aarhus C, Denmark}
\date{\today}

\maketitle

\begin{abstract}
The complex energies of the three-body resonances for one infinitely
heavy particle and two non-interacting light particles are the sum of
the two contributing two-body complex resonance energies.  The bound
state of a Borromean system originates from a resonance when the third
interaction is introduced, a finite mass is allowed and proper angular
momentum coupling is included. The relative importance of these
contributions are investigated and the resulting structure of
Borromean systems are traced back to the two-body continuum
properties. The $0^+$ and $2^+$ states in $^{6}$He result from
neutron-core $p$-states and the ground and first excited state of
$^{11}$Li originate from neutron-core $s^2$ and $sp$-states.
\end{abstract}
\end{frontmatter}

\par\leavevmode\hbox {\it PACS:\ } 21.45.+v, 31.15.Ja, 25.70.Ef

\section{Introduction.}

Three particles can form a bound state even when none of the two-body
subsystems are bound \cite{nie01}. The attraction responsible for such
Borromean states is almost inevitably of short range.  The existence
was highlighted in the measurements of the unexpected large reaction
cross section of $^{11}$Li implying a very large radius \cite{tan01}.
Subsequently halos were used to denote such systems \cite{han95} with
simple scaling properties but independent of the interactions
\cite{jen04}.   Necessary conditions for halo occurrence are
small binding energy, low relative angular momentum and a vanishing or 
relatively weak repulsive Coulomb interaction.

Borromean systems are by definition weakly bound, since the three-body
binding eventually originates from two-body interactions which must be
too weak to bind the two-body systems.  The stability of Borromean
systems is surprising because a realistic system of a comparatively
heavy core surrounded by two mutually non-interacting particles has
three-body resonance energies equal to the sum of two particle-core
resonance energies \cite{gar03}.  Thus if none are bound the
three-body system is also unbound and the Borromean system does not
exist.

However, three different effects can independently change this
conclusion, i.e. a particle-particle attraction, a finite core-mass
allowing favorable angular momentum coupling, and the smaller kinetic
energy in the three-body system compared to the three two-body
systems.  The purpose of the present letter is to estimate these three
effects and map out the road from the two-body continuum to Borromean
states. We shall concentrate on systems where one particle mass is 
substantially larger than the other two. Then our initial schematic 
structure of one infintely heavy mass and two non-interacting particles 
still leaves strong fingerprints in the final system with realistic masses 
and interactions. Systems with comparable masses can also be studied in 
the same way but now the realistic system can have rather little in common 
with the initial schematic configuration. 

In the quantitative description we shall use 
universal terms suitable for understanding the generic origin of Borromean 
systems in general and in particular the well studied nuclear examples of
$^{6}$He and $^{11}$Li. A welcome side effect is that possible
three-body (continuum and bound-state) structures become easier to
predict directly from properties of the two-body interactions.

\section{The foundation.}

We label the three particles by $i$=1,2,3 and denote their masses and
two-body interactions by $m_i$ and $V_{ij}$. All three interactions
are too weak to support a bound state. The two-body energies of the
virtual states and resonances are $E^{(n)}_{ij}$. For the schematic
system, where $V_{23}=0$ and $m_1$=$\infty$, the three-body system has
corresponding resonance and virtual state energies $E$=$E^{(n)}_{12}$+
$E^{(m)}_{13}$. This addition theorem holds for all angular momenta
compatible with the usual coupling rules. The energy is degenerate if
more than one total angular momentum is allowed. These observations
follow from the separability of the two subsystems \cite{gar03}.

The procedure is now to relax the stringent conditions of the
schematic model until a realistic Borromean system is reached. It is
then convenient to maintain the two-body properties of the two
initially interacting two-body systems. This is achieved with
unchanged potentials $V_{12}$, $V_{13}$ and reduced masses $\mu_{12}$,
$\mu_{13}$ (vary $m_2$ and $m_3$) as $m_1$ becomes
finite. The three-body resonances can then be studied as function of
$m_1$ for different partial waves, angular momenta, parities,
symmetries, and for non-vanishing $V_{23}$.  Conditions for the
appearance of Borromean systems must then emerge.

For spin independent two-body interactions the three-body
wave function factorizes into coordinate and spin parts. The energies
of the three-body resonances and bound states are then independent of
the particle spins and their couplings. This is still valid when
particles 2 and 3 are identical fermions or bosons. The spin part of
the wave function can always be used to establish the correct
(anti)symmetry under exchange of particles 2 and 3. Therefore the
orbital angular momentum part of the wave function is the same as for
non-identical spin-zero particles 2 and 3.  Furthermore, due to
separability for $m_1=\infty$ and $V_{23}=0$, the three-body states
with different orbital angular momentum $L$ are degenerate even if
particles 2 and 3 are identical.

The only exception is when particles 2 and 3 are identical bosons with
zero spin.  Then the spin part in the wave function is not available
to establish the symmetry under exchange of the particles, and the odd
values of the relative orbital angular momentum between the two bosons are
therefore strictly forbidden. This has consequences even when
$V_{23}=0$. For instance, if the core-boson interaction only has a
$p$-wave, the minimum value of the hypermomentum for the even parity
states should normally be $K_{min}$=2. However, to get $L=1$, the
smallest values of the relative orbital angular momenta are $2$ both
between the bosons and between their center of mass and the core. This
implies that $K_{min}$=4.  The effective radial potentials therefore
vary strongly from $L=1$ with $K_{min}=4$ to $L=0,2$ where
$K_{min}=2$, see \cite{nie01}.

We need a reliable method to compute three-body resonances and bound
states.  We choose the Faddeev hyperspherical adiabatic expansion
method \cite{nie01,jen97} combined with the complex scaling method
\cite{gar03,ho83,myo01}. For this we define the only length
coordinate, the hyperradius $\rho$, as
\begin{equation} \label{e20}
 m \rho^2 \equiv  \frac{1}{ M}  \sum_{i<k} m_i m_k 
(\bm{r}_i - \bm{r}_k)^2  \;  ,
\end{equation}
where $\bm{r}_i$ is the coordinate of particle $i$, $M = \sum m_i$ and
$m$ is an arbitrary normalization mass. The complex scaling then
amounts to the substitution $\rho$$\rightarrow$$\rho \exp({i\theta})$
in the Faddeev equations.  The wave function of a resonance then falls
off exponentially when the angle $\theta$ is larger than
$\theta_r$$\equiv$$\arctan(\kappa_I/\kappa_R)$, i.e. the angle
corresponding to the complex momentum
$\kappa$=$\kappa_R-i\kappa_I$=$\sqrt{2mE/\hbar^2}$ of the complex
resonance energy $E$. The asymptotic large $\rho$ behavior of the
different radial wave functions $f_{nn'}$ are \cite{nie01,gar03}
\begin{equation} \label{eq5}
f_{nn'} 
\rightarrow \sqrt{\rho} H_{K+2}^{(1)}(|\kappa| \rho e^{i(\theta-\theta_R)}),
\end{equation}
where $K$ is the integer (hypermomentum) related to the asymptotic adiabatic 
radial potential and $H_{K+2}^{(1)}$ is the outgoing Hankel function of first
kind.  Bound states with the true real and negative energies are also
eigensolutions to the complex rotated equations.

\section{P-waves: $^6$He}

\begin{figure}
\begin{center}
\vspace*{-2cm}
\epsfig{file=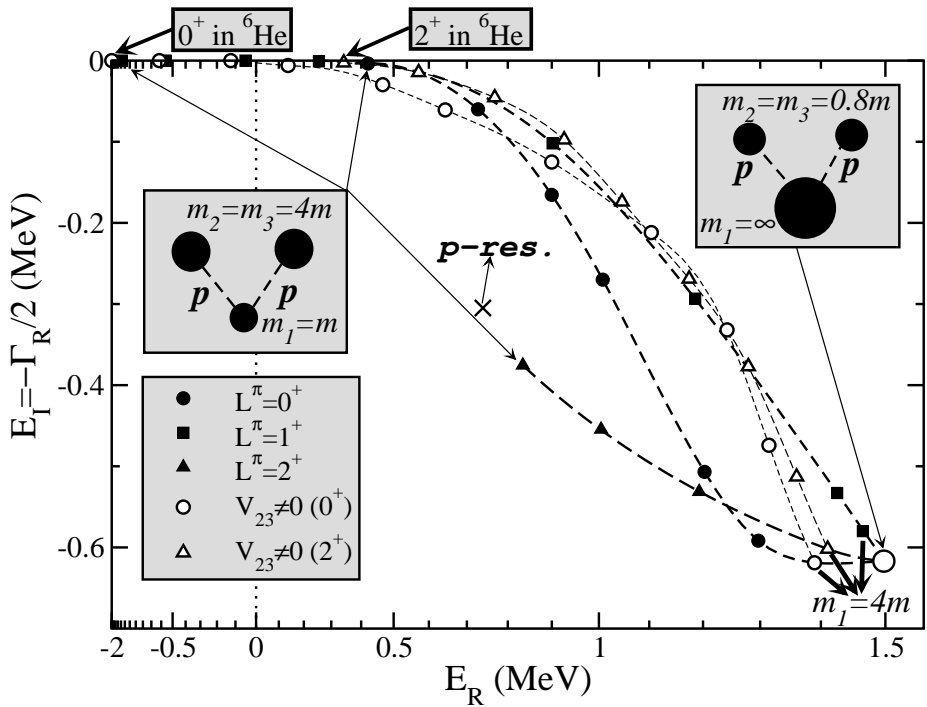,scale=0.5,angle=-90}
\end{center}
\caption[]{Real and imaginary values of resonance energies for a three-body 
system with reduced masses $\mu_{12}=\mu_{13}=0.8m$ ($m$ is the
nucleon mass).  The cross indicates the position of the $p$-resonance
for the 1-2 and 1-3 two-body systems.  The big open circle indicates
the three-body resonance for $m_1=\infty$ and $V_{23}=0$.  The solid
circles, solid squares, and solid triangles show the energies of the
0$^+$, 1$^+$, and 2$^+$ resonances as function of $m_1$.  The open
circles and open triangles show the energies of the 0$^+$ and 2$^+$
resonances for $m_1=4m$ when $V_{23}$ change from zero to full
neutron-neutron strength. }
\label{fig1}
\end{figure}

Let us consider an infinitely heavy core ($m_1=\infty$) and two
non-interacting identical particles 2 and 3 ($V_{23}=0$) such that
$\mu_{12}$=$\mu_{13}$=0.8$m$, where $m$ is the nucleon mass. The
core-particle interaction is a spin-independent gaussian acting only
on $p$ waves with a two-body resonance energy and width ($E_R$,
$\Gamma_R$)=(0.74,0.61) MeV indicated by the cross in Fig.\ref{fig1}.
These parameters correspond to the lowest $p_{3/2}$-resonance in
$^5$He.  The three-body system has a resonance with twice the energy
and width, i.e.  $(E_R,\Gamma)$=(1.48,1.22) MeV indicated by a big
open circle in Fig.\ref{fig1}. The degeneracy is 3 corresponding to
the orbital angular momenta $L^\pi$=0$^+$,1$^+$,2$^+$ except for two
identical bosons with spin zero where $L^\pi=1^+$ is excluded for
symmetry reasons.

In Fig.\ref{fig1} we show how the resonances of $L^\pi$=0$^+$, 1$^+$,
and 2$^+$ vary as $m_1$ changes from $\infty$ to $m$ (solid symbols). 
The degeneracy
is broken and all three states move towards more binding resulting
entirely from the changing contribution of the center of mass
kinetic energy.  The $L$=1 three-body state is even bound for
$m_1$=$m$.  Thus a Borromean state can exist even if two of the particles
do not interact.

Along these three curves the points indicated by the thick arrows in
Fig.\ref{fig1} correspond to $m_1$=4$m$ and $m_2$=$m_3$=$m$. This
three-body system is similar to $^6$He, except that the interaction
$V_{23}$ is equal to zero.  We now introduce the realistic
spin-dependent neutron-neutron interaction $V_{23}$ multiplied by a
factor varying from 0 to 1.  The 0$^+$ and $2^+$ states then move from
the close-lying $m_1=4m$ points to respectively binding of 2 MeV
(Borromean state) and the resonance position $(E_R,\Gamma_R)=$(0.34
MeV,0.01 MeV).  The effect of changing from  $m_1=\infty$ to
$m_1=4m$ is small, and therefore the effect responsible for the existence 
of the $0^+$ Borromean state and the low-lying 2$^+$ resonance is the
interaction between the neutrons.

The computed 0$^+$ and 2$^+$ states in the schematic model are more
bound than the experimental values for the corresponding states in
$^6$He.  This overbinding has two main sources.  First, the
core-neutron spin-orbit interaction is neglected and the $p_{1/2}$
resonance appears at 0.74 MeV instead of the $2$ MeV measured for
$^5$He. Second, we neglected the highly repulsive alpha-neutron
$s$-wave interaction which accounts for the Pauli principle by
excluding overlap to the neutron $s$-state within the alpha particle.
A detailed calculation taking these facts into account is available in
\cite{gar97,fed03}.

The neutron-neutron interaction is predominantly of $s$-wave
character, and an $L=1$ state is therefore only possible with a
relative $p$-state between core and center of mass of the two
neutrons. Then the neutron-neutron interaction has no effect on the
1$^+$-state which remains at the initial $m_1=4m$ point.  Furthermore,
the 1$^-$ state does not appear numerically as a resonance, because
one virtual $s$-state and one $p$-resonance produce an $S$-matrix pole
on the unphysical sheet.  We have not found any 1$^-$ resonance even
for a realistic alpha-neutron interaction with a repulsive partial
$s$-wave.

\section{Mixed S and P-waves: $^{11}$Li.}

\begin{figure}
\begin{center}
\vspace*{-2cm}
\epsfig{file=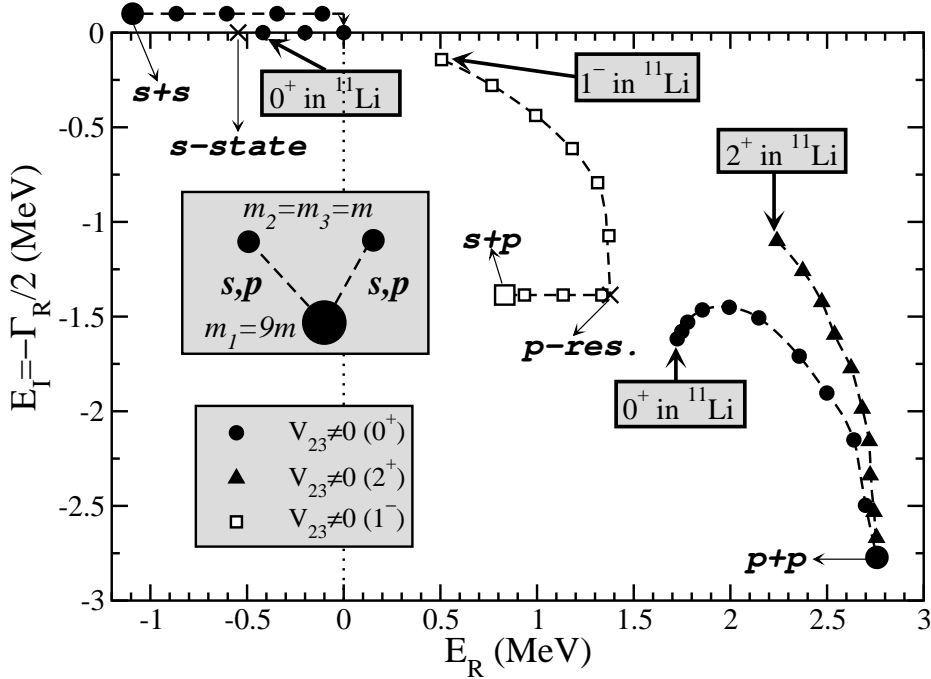,scale=0.5,angle=-90}
\end{center}
\caption[]{Real and imaginary values of resonance energies for a three-body 
system with reduced masses $\mu_{12}=\mu_{13}=0.9m$ ($m$ is the
nucleon mass).  The crosses indicate the positions of virtual
$s$-state and $p$-resonance for the 1-2 and 1-3 two-body systems.  The
big solid circles and big open square indicate the three-body poles for 
$V_{23}=0$.  The solid circles, solid triangles, and open squares show the 
energies of the 0$^+$, 2$^+$, and 1$^-$ states for $m_1=9m$ when 
$V_{23}$ becomes finite.}
\label{fig2}
\end{figure}

We now extend to include both $s$ and $p$-waves.  Particles 2 and
3 are again identical particles, the interactions $V_{12}=V_{13}$ with
reduced masses $\mu_{12}=\mu_{13}=0.9m$ produce a virtual $s$-state at
$-0.54$ MeV and a $p$-resonance with ($E_R, \Gamma_R$)=(1.38, 2.77)
MeV (crosses in Fig.\ref{fig2}). For $V_{23}=0$ the computations in practice 
correspond to
$m_1=9m$, $m_2=m_3=m$ which for $p$-waves alone leads to resonances
similar to the results for $m_1=\infty$, see Fig.\ref{fig1}. As in 
the previous section, for pure $p$-waves, the $L^\pi=0^+,1^+,2^+$ three-body 
resonances are given by twice the energy of the two-body $p$-resonance
(big solid circle labeled as $p+p$ in Fig.\ref{fig2}).

In addition there is a second 0$^+$ virtual state with only $s$-wave
components, and a $1^-$ pole with mixed two-body $s$ and
$p$-components. Both these $S$-matrix poles appear on the unphysical
Riemann sheet at twice the two-body $s$-state energy and at the sum of $s$
and $p$ energies, respectively (big solid circle labeled as $s+s$ and 
big open square in Fig.\ref{fig2}).  They cannot be found numerically with
the complex scaling method \cite{fed03}, but a sufficiently strong
attractive interaction $V_{23}$ moves the pole onto the physical sheet
where they appear as  resonances \cite{gar03}.

We choose again $V_{23}$ as the realistic neutron-neutron interaction
multiplied by a factor varying from 0 to 1. The two identical fermions
with spin 1/2 suppress the 1$^+$.  The evolution with the strength  of
$V_{23}$ is shown in Fig.\ref{fig2}. The 0$^+$ and the 2$^+$ states,
originating exclusively from the $p$-components, both move towards more
binding arriving at ($E_R,\Gamma_R$)=(1.73,3.24) MeV and (2.24,2.20)
MeV for the 0$^+$ and 2$^+$ states, respectively. For full
neutron-neutron strength the core-neutron $p$-wave components in the
0$^+$ state correspond to a probability of about 80\% whereas the 2$^+$
state entirely must be 100\% $p$-waves.  The 2$^+$ state has a smaller
width and a larger energy than the 0$^+$ state, but they are both far
from being bound states.

The other 0$^+$ state, originating exclusively from two-body
$s$-components, is initially a virtual three-body pole on the negative
energy axis of the unphysical Riemann sheet. As the attraction
$V_{23}$ is switched on this pole moves towards the origin and continues
smoothly onto the physical sheet and becomes a Borromean bound state
of energy $E_R=-0.42$ MeV. The core-neutron $p$-wave components in
this state is slightly above a probability of 30\%.

The 1$^-$ state originating from mixed $s$ and $p$ two-body states is
also initially on the unphysical sheet, but as $V_{23}$ is continuously 
switched on the pole appears at some point as a resonance with the
complex energy of the two-body $p$-resonance \cite{gar03}.  This
evolution is shown in Fig.\ref{fig2} with the arrival for full
neutron-neutron interaction at ($E_R,\Gamma_R$)=(0.51,0.29) MeV.  The
core-neutron components remain for parity reasons a mixture of
$s$ and $p$-states.

The spectrum obtained with the full neutron-neutron interaction is the 
one of $^{11}$Li with the assumption of zero core spin. It is known 
experimentally that $^{11}$Li  has a
Borromean 0$^+$ bound state at $-0.3$ MeV with a $p$-wave content of
40\%-50\%, and also evidences for a 1$^-$ resonance at about 
1 MeV \cite{kor96} have been found.  There is not experimental information 
about other resonances, but different calculations have predicted additional
excited states \cite{aoy02}. 

The $^{11}$Li calculations shown in Fig.~\ref{fig2} have unrealistic
features.  The core-neutron interaction should be more attractive to
reproduce the experimental constraints of an $s$-state below 100 keV
and a $p$-resonance at around 0.5 MeV.  The overbinding mentioned in
connection with $^6$He is also a problem here due to neglect of
spin-orbit potential and the simplified treatment of the Pauli
principle.  However, a conceptually different problem is the spin
dependence of the core-neutron interaction due to the finite spin of
the $^{11}$Li core. The consequence is an unavoidable spin-splitting
lifting the degeneracy from the beginning and furthermore lowering the
crucial two-body virtual state and resonance energies.  All these
effects, partially canceling each other, conspire to make the
$^{11}$Li system complicated.

Still the interactions used in Fig.\ref{fig2} provide a realistic
illustration of the relative sizes of the principal effects.  In
particular the ground state originates from the combination of two
$s$-components, and the first rather low-lying 1$^-$ excited state
originates from the mixing of $sp$-components. Also, additional
excited states (0$^+$ and 2$^+$) arise from the two $p$-components.
In fact these four states are consistent with the spectrum found for a 
system like $^{12}$Be, that it is known experimentally to have a bound 
0$^+$ state, a 0$^+$ resonance, a 1$^-$ resonance, and a 
2$^+$ resonance \cite{iwa00,shi03}.

\section{Realistic $^{11}$Li: Finite core spin.}

\begin{figure}
\begin{center}
\vspace*{-2cm}
\epsfig{file=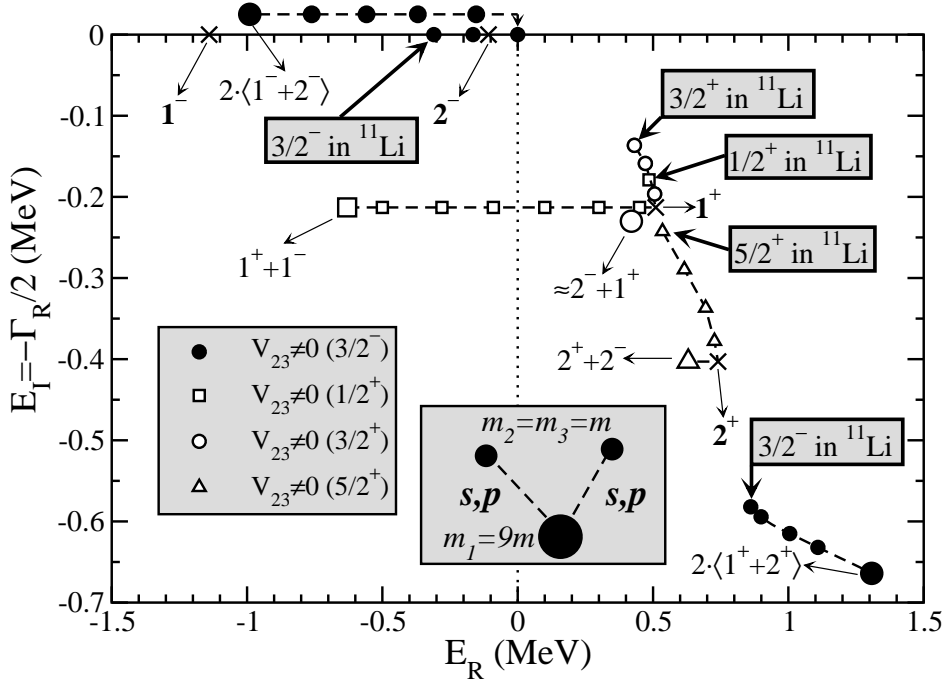,scale=0.5,angle=-90}
\end{center}
\caption[]{Real and imaginary values of resonance energies for $^{11}$Li with
finite core spin of 3/2. The reduced masses are $\mu_{12}=\mu_{13}=0.9m$. 
The two-body systems 1-2 and 1-3 have virtual $s$-states and a $p$-resonances 
at the energies indicated by the crosses. The big symbols are the three-body 
states of the system for $V_{23}=0$.  The solid circles, open squares, open 
circles, and open triangles show the energies of the 3/2$^-$, 1/2$^+$, 
3/2$^+$, and 5/2$^+$ states when the neutron-neutron interaction is 
progressively increased. }
\label{fig3}
\end{figure}

The immediate consequence of a non-zero core spin $s_1$ is the
increase in the number of low-lying core-neutron resonances. Every
core-neutron level of angular momentum $j_2$ is split into a series of
states with total angular momentum from $|j_2-s_1|$ to $j_2+s_1$.  A
realistic model should then reproduce experimental information of (the
most important of) the degeneracies and complex energies of these
states.  Then the neutron-core interaction necessarily contains
spin-spin and spin-orbit operators which must be chosen judiciously.
This is particularly important when the particles surrounding the core
are fermions identical to those within the core. Consistency between
the antisymmetrization and the spin operators is indispensable to
avoid catastrophic results for the three-body system \cite{gar03b}.

To trace the origin of such Borromean systems exemplified by $^{11}$Li
with core-spin and parity $3/2^-$ we again turn to the schematic model
where $m_1=\infty$ and $V_{23}$=0.  The three-body hamiltonian $H$ is
still a sum of two two-body hamiltonians $H_{12}$ and $H_{13}$ with
eigenvalues $E_{12}^{(j_{12})}$ and $E_{13}^{(j_{13})}$, respectively.
The states are labeled by the two-body angular momenta $j_{12}$ and
$j_{13}$ obtained by coupling the spin $s_1$ of the core and the total
two-body angular momenta $j_2$ and $j_3$ of the particles in their
motion relative to the core.

The relevant $^{9}$Li-neutron states are the $s_{1/2}$ and $p_{1/2}$
partial waves which, combined with spin and parity of the core, result
in the two $^{10}$Li doublets 1$^-$/2$^-$ and 1$^+$/2$^+$.  We choose these
$^{10}$Li continuum states at $-0.10$ MeV (2$^-$), $-1.14$ MeV
(1$^-$), 0.51-$i$0.21 MeV (1$^+$), and 0.74-$i$0.40 MeV (2$^+$).
These energies (crosses in Fig.\ref{fig3}) are slightly above the ones 
suggested by the
experiments and by more sophisticated calculations \cite{gar02}. This
allows us to reproduce the experimental three-body energy with the
present model.

When only the $s_{1/2}$
and $p_{1/2}$ waves are considered in $^{10}$Li, the possible
three-body $^{11}$Li states are characterized by total
angular momentum and parity $j^{\pi} = 1/2^{\pm}, 3/2^\pm, 5/2^{\pm}$. 
The ground state has
$j^{\pi} = 3/2^{-}$ and the positive parity states correspond to the 
$1^-$-excitations. These three-body states are antisymmetric under neutron 
interchange.  In analogy to the measured structure of the 
$^{12}$Be spectrum \cite{shi03} the $s_{1/2}$ and $p_{1/2}$ states inevitably
lead to two 3/2$^-$ states. 

In the schematic model the energies of the two $j^{\pi} =
3/2^{-}$-states are twice the average energy of the doublets
corresponding to two neutrons either in the 1$^-$/2$^-$ (100 \% $s^2$)
or in the 1$^+$/2$^+$ (100 \% $p^2$) states.  These average energies
are shown in Fig.\ref{fig3} as big solid circles at $-0.99$ MeV and
$1.31-i0.66$ MeV.  The correct finite core mass only moves these points
very little.  When the neutron-neutron interaction is progressively 
switched on the first $3/2^-$-state moves along the negative energy
axis until the system becomes bound and at full strength the Borromean
state is found at $-0.30$ MeV with the $s^2$ content reduced to 70\%.
The second $3/2^-$-state also moves and reaches for full strength the
resonance position $(E_R,\Gamma_R)$=(0.86,1.16) MeV with the $p^2$
content reduced to 93\%.

The structure of the $1^-$-excitations must correspond to one neutron
in one of the two $s$-states in $^{10}$Li, and the other neutron in
one of the $p$-states.  For the $1/2^+$ and the $5/2^+$ states, the
two neutrons necessarily occupy the $1^-$/1$^+$ and 2$^-$/$2^+$
states, respectively.  Therefore, in the schematic model the
three-body energies are $E_{12}^{(1)}+E_{13}^{(1)}$ and
$E_{12}^{(2)}+E_{13}^{(2)}$ as shown in Fig.\ref{fig3} by a big open
square and a big open triangle.  These states are not three-body
resonances, but virtual $s$-states with respect to the two-body
resonance.  When the neutron-neutron interaction increases, at some
point these poles on the unphysical Riemann sheet appear as three-body
resonances through the two-body resonance energies indicated by the small
open squares and the small open triangles.

Finally, in the schematic model two $3/2^+$-states are found
at $0.42-i0.23$ MeV and $-0.42-i0.38$ MeV, which approximately
correspond to the neutrons in the $2^-$/1$^+$ and the
$1^-$/$2^+$-states, respectively. Only the first of these states shows
up as a three-body resonance when the neutron-neutron interaction is
switched on. The other pole remains on the unphysical Riemann sheet.
All the three 1$^-$ resonances move towards the origin reaching an energy of
about 0.5~MeV and widths of 0.3-0.5~MeV. Evidence for a combination
of these overlapping resonances is available \cite{kor96}.

The degeneracy of these $1^-$-excitations is broken by the core-spin
dependence of the $^{9}$Li-neutron interaction.  The resonance
positions are very sensitive to this spin dependence which must be
chosen consistent with the Pauli principle and the mean-field
treatment of the neutrons within the core.  Violation of these
constraints can rather easily lead to very wrong energies of the
$1^-$-resonances or even to bound states.

\section{Conclusions.}

We have traced the origin of Borromean systems to the properties of
the two-body subsystems. One infinitely heavy core mass and two
non-interacting light particles is a separable system where the
complex energies of the $S$-matrix poles are obtained by adding
the corresponding two-body energies. The three-body states then lie higher
than the individual two-body states.  The existence of Borromean bound
states means that attractive effects must overcompensate for this
initial loss of energy in the three-body system. Three contributions
are responsible, i.e.  the finite core mass reduces the positive
kinetic energy, the angular momentum couplings favor special
structures, and the light particle interactions are attractive in specific
partial waves.

We illustrate with established Borromean nuclei where two neutrons
surround the core.  Then only the short-range interaction
contributes. Inclusion of the Coulomb repulsion destroys the initial
assumption of separability for an infinitely heavy core.  The poles
would move towards less binding and the widths would be much smaller
due to the repulsive and confining effects of the Coulomb
barrier. Otherwise we expect the same qualitative behavior.

We conclude that Borromean systems can arise separately from each of
the above three contributions. With known energies of low-lying
virtual states and resonances we can predict both the relative
importance of the three effects and the properties of the resulting
three-body ground and excited states.  These investigations have only
recently become possible with the implementation of the powerful
techniques to handle both discrete and continuum three-body states.

\end{document}